\documentclass[twoside]{article}
\usepackage{fleqn,espcrc2}
\usepackage{graphicx}

\newcommand{\AmS}{{\protect\the\textfont2
  A\kern-.1667em\lower.5ex\hbox{M}\kern-.125emS}}
\newcommand{\imag}{\hbox{Im}\,}
\newcommand{\real}{\hbox{Re}\,}
\newcommand{\pepe}{\hbox{P.P.}\,}
\newcommand{\dd}{\hbox{d}\,}

\title{In memory of Paco Yndur\'ain: A precise determination of $\pi\pi$ scattering from experiment and dispersion relations}

\author{
J.R.Pel\'aez\address{
Dept. de F\'{\i}sica Te\'orica II, Univ. Complutense de Madrid.
28040 Madrid. Spain}, R. Garc\'{\i}a Mart\'{\i}n$^a$,
R.~Kami\'nski\address{Department of Theoretical Physics
Henryk Niewodnicza\'nski Institute of Nuclear Physics,
Polish Academy of Sciences,
31-342, Krak\'ow, Poland.}
and F. J. 
Yndur\'ain\address{
Dept. de F\'{\i}sica Te\'orica, C-XI
 Univ. Aut\'onoma de Madrid,
 Canto Blanco,
E-28049, Madrid, Spain.}
}
       
\begin{document}

\begin{abstract}
This talk is dedicated to the memory of 
Paco Yndur\'ain, the original speaker in the conference. 
After a short account of his scientific career,
we briefly review our ongoing collaboration to determine precisely the
$\pi\pi$ scattering amplitude including the most recent data
by means of Forward Dispersion Relations and Roy Equations.
A remarkable improvement in precision over the intermediate energy region
is obtained by using once-subtracted Roy Equations in addition to the standard
twice-subtracted ones.
\end{abstract}

\maketitle

\section{ Paco Yndura\'in, in memoriam}

F. J. Yndur\'ain, ``Paco'' for his friends, was very dear for 
this QCD96-QCD08 Conference series, since, apart from an 
 active participant, 
he was member of the Advisory Committee in
all editions.
For the present QCD08, he was to review our work, but, 
very unfortunately, he passed away
just before the beginning of the Conference. 
We thank prof. S. Narison for offering us the chance to
give the same talk together with a brief account of Prof. Yndur\'ain
 outstanding scientific career.

In the University of Zaragoza, Spain, Paco obtained his Master 
degree in Mathematics in 1962 and defended  in 1964 his PhD 
thesis in Physics, ``Definitions of Hamiltonians and Renormalization'',
supervised by A. Galindo. 
In Zaragoza he also held an assistant professorship (1964/66), 
before moving to 
New York University as a Fulbright Fellow (1966/67) 
and Associate Researcher (1967/68). He then became 
Research Fellow at CERN (1968/70),
an institution to which he was specially attached, being senior 
scientific Associate in 1976 and 1985, and elected Member 
of the Scientific Policy Committee from 1988 to 1994. 

Finally, after a short period in the 
Complutense University of Madrid in 1970, 
he became
full professor in the Aut\'onoma Univesity of Madrid, 
where he had a decisive 
influence in shaping one of the best Theoretical Physics 
Departments and Particle Physics Groups in Spain. Actually, 
he was Director of the Department twice (1974/77-1981/84), dean of the Science Faculty 
in 1975 and research vice-rector (1978/81).

Paco's intense activity was rewarded with many friends 
and collaborations all over the world, becoming visiting
researcher at CERN, Michigan, Marseille, 
NIKHEF, Brookhaven, Orsay, Saclay, Viena, La Plata, Bogot\'a, Caracas...
He was also founding member of the European Physical Society, 
invited member of the American Association for the Advancement 
of Science, the New York Academy of Science and both the 
Mathematical and Physics Spanish Royal Societies.
In addition, he became elected member of the European Board of 
High Energy Physics of the European Physical Society (1983/89), 
and elected president of the Physics and Chemistry section of the 
Spanish Royal Academy in 2002. But his influence extended beyond Physics
as scientific advisor of IBM (1983/85),
of the Institute for Scientific Research of Kuwait in (1980/82),
and as elected member, by the Spanish Senate, of the 
Spanish University Council, as well as elected member of the 
World Innovation Foundation in London (2001) and the {\it Academia Europaea}.

Paco was awarded the gold medal of the Spanish Royal Academy 
and was an honorific collaborator of ICTP in Trieste, 
La Plata University, and Cavaliere Ufficiale nell'Ordine al 
Merito della Reppublica Italiana, a country very dear for him.

\vspace{9pt}
\includegraphics[width=7cm]{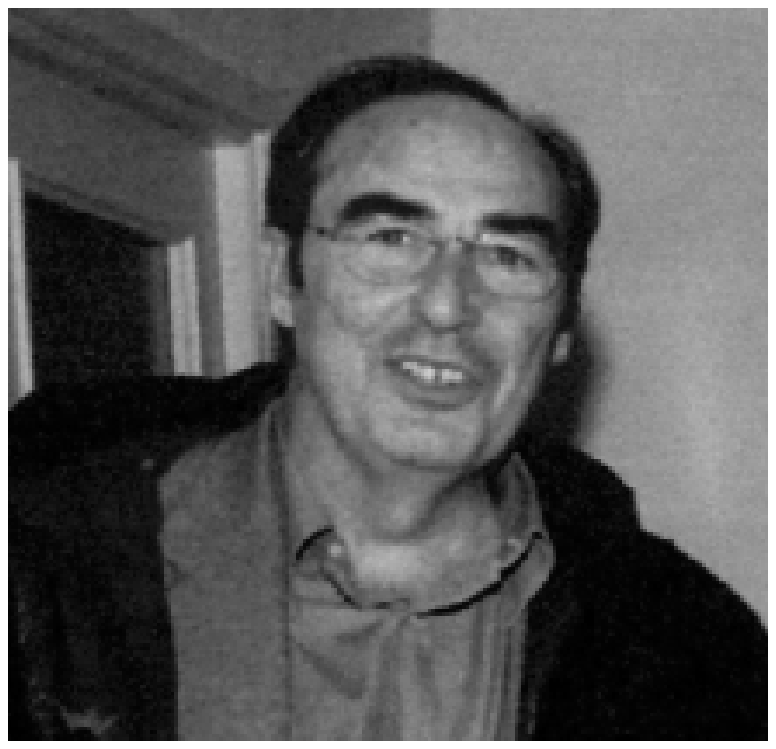}

\subsection{Paco, Particle Physics and QCD}

Paco authored or coauthored more 
than 160 publications, including articles and talks.
Unfortunately in this space we can only highlight a few. 
In his New York period 
he was interested in analyticity methods for scattering theory, that he 
gathered in
a Review of Modern Physics in 1972 \cite{Yndurain:1972ix}. Around that
time, QCD, the subject of this conference series,
 was established as the theory of strong interactions. This is where
Paco obtained his most celebrated results, 
like the  calculations of structure functions in the 70's and 80's 
\cite{Yndurain:1977wz}, 
the hadronic contributions to the muon g-2 in the 80's and 00's \cite{Casas:1985yw}, 
quarkonium calculations in the 90's \cite{Titard:1993nn}, as well as 
light quark and chiral symmetry QCD calculations in the early 80's \cite{Becchi:1980vz}.

The participants in this QCD08 conference might be more familiar
with his QCD work, but he also made other well known 
contributions to the Standard
Model phenomenology, like the calculation of radiative corrections 
to $WW$ scattering \cite{Veltman:1989vw}. He also made some well known phenomenology 
beyond the Standard Model
\cite{Jarlskog:1978uu}  but always limited within his
very cautious and critical attitude towards Standard Model extensions.

In recent years, motivated by the very 
recent and precise experimental results,
he was looking back 
to some of his initial interests, about analytic properties of 
pion-pion scattering amplitudes \cite{Kaminski:2006qe,Yndurain:2007qm}, the subject of this talk.

\subsection{Books}

Paco dedicated an enormous effort and enthusiasm to communicate Physics 
and Science. Of special relevance for the participants 
in this QCD Conference is his famous book -- a classic -- 
``QCD: An introduction to the Theory of Quarks and Gluons'' (Springer-Verlag 1983).
For many of us in the young -- well, maybe not so young -- generations 
of Particle Physicists, Paco's book has been THE BOOK where to learn QCD. 
He kept improving it throughout four different 
editions (from the second it is
entitled ``The Theory of Quark and Gluon Interactions'').
More recently he also wrote a book on ``Relativistic Quantum Mechanics: and introduction
to Field Theory'' (Springer-Verlag 1996), and another one on 
``Mec\'anica Cu\'antica'' (In Spanish, 2nd Ed. Ariel S.A. 2003).

Less known to the international community -- they are written in Spanish--
are his  books on Popular Science and ``Speculation'', as he called them, 
whose translated titles are:
``Unified Theories and the constituents of matter'',
``Who walks out there'',
``Electrons neutrinos and quarks'' and 
``The Challenges of Science''.

\subsection{Why the name ``Paco'' and farewell}

Anyone called ``Francisco'' in Spain can also be called ``Paco''. The reason 
is that St. Francesco de Assisi -- San Francisco 
in Spanish--, was the founder of Franciscan monastic community
and so he was the ``Father of the Community'' 
or {\it ``Pater Comunalis''} in latin. Abbreviated: {\it Pa.Co.}

But in the case of F.J. Yndur\'ain, ``Paco'' was 
much more than just a nickname. In dire times for 
Spanish Scientific research, he was able to produce first level results and 
was able to gather a research group on Particle and Theoretical Physics
that later on became the Theoretical Physics 
Department of the Aut\'onoma University in Madrid, extending his influence and 
scientific attitude much beyond.
With his career and example he really became one of the {\it ``PAter COmunalis''} 
of the Spanish Particle Physics Community.

Paco, we will miss you as a friend, but also your enthusiastic and contagious 
vitality not only about Physics but life as a whole. 
{\it Sit tibi terra levis} --- 
may the earth rest lightly on you.

Of course, knowing Paco, I guess that by now he would be highly embarrassed 
and yelling from the back... Enough is enough! Let's have fun with Physics!!... -- 
Let's not make him wait any longer.

\section{Dispersive approach to $\pi\pi$ scattering}

A precise determination of the $\pi\pi$ scattering amplitude
at low energies
is relevant for the study of Chiral Perturbation Theory (ChPT),
quark masses, the chiral condensate \cite{Gasser:1983yg} 
and, at intermediate 
energies, for the properties of the controversial sigma meson.
However, the existing experimental information from $\pi\pi$ 
scattering has many conflicting data sets at 
intermediate energies and for many years very little data 
in the low energy region close to threshold. The interest in the topic has
been renewed with recent \cite{Batley:2007zz}
and precise experiments
on kaon decays, relatively easy to relate to $\pi\pi$ scattering.

From the theoretical side, this process is very special due to 
the strong constraints from  isospin, crossing and chiral symmetries, but mostly
from analyticity. The latter allows for a very rigorous
dispersive integral formalism that
relates the $\pi\pi$ amplitude 
at any energy with an integral over the whole energy range, increasing the precision 
and providing information
on the amplitude even at energies where data are poor,
or in the complex plane. Remarkably, it is model independent since
it makes the data parametrization irrelevant once it is included in the integral. 
The dispersive approach is thus well suited to
study the threshold region or  poles in the complex plane associated to resonances.
Our recent works \cite{Kaminski:2006qe,Yndurain:2007qm} make use of two complementary 
dispersive approaches, in brief:

$\bullet$ {\it Forward Dispersion Relations (FDRs):} 
Calculated at $t=0$ so that the 
amplitude unknown large-t behavior is not needed.  
There are two symmetric and one asymmetric isospin combinations, 
to cover the isospin basis.  The symmetric ones,
$\pi^0\pi^+$ and $\pi^0\pi^0$, have two subtractions
\begin{eqnarray}
&\hspace*{-3.5cm}\real F(s,0)-F(4M_{\pi}^2,0)=
\label{FDR1} \\
&
\frac{s(s-4M^2_\pi)}{\pi}\,\pepe\int_{4M_{\pi}^2}^\infty
\frac{(2s'-4M^2_\pi)\,\imag F(s',0)\,\dd s'}{s'(s'-s)(s'-4M_{\pi}^2)(s'+s-4M_{\pi}^2)}
\nonumber
\end{eqnarray}
where $F$ stands for the
$F_{0+}(s,t)$ or $F_{00}(s,t)$ amplitudes. They are very precise since
all the integrand contributions 
 are positive.
The antisymmetric isospin combination $I_t=1$ reads
\begin{eqnarray} 
&\hspace*{-4cm}F^{(I_t=1)}(s,0)=
\label{FDR2} \\
&\frac{2s-4M^2_\pi}{\pi}\,\pepe\int_{4M^2_\pi}^\infty\dd s'\,
\frac{\imag F^{(I_t=1)}(s',0)}{(s'-s)(s'+s-4M^2_\pi)}. 
\nonumber
\end{eqnarray}
All FDRs are calculated up to $\sqrt{s}\simeq1420$~MeV

$\bullet$ {\it Twice subtracted Roy Equations (RE):} they are an infinite set of coupled equations \cite{Roy:1971tc}, equivalent to nonforward 
dispersion relations plus some $t-s$ crossing symmetry.
They are well suited to study poles of resonances since they are 
 written in terms of partial waves $f^{(I)}_l$ 
of definite isospin I and angular momentum $l$.
The complicated left cut contribution is rewritten 
as a series of integrals over the physical region:
\begin{eqnarray}
&&\hspace{-.5cm}\real f^{(I)}_l(s)=C_l^{(I)}a_0^{(0)}+{C'_l}^{(I)}a_0^{(2)}
\label{Roy}\\
&&+\sum_{l',I'}
\pepe\int_{4M^2_\pi}^\infty\dd s' 
K_{l,l';I,I'}(s',s)\imag f^{(I')}_{l'}(s'),
\nonumber
\end{eqnarray}
where the $C_l^{(I)}$, ${C'_l}^{(I)}$ constants and $K_{l,l';I,I'}$ kernels
are known.  The calculation is truncated at $l<2$ and at some cutoff
energy $s_0$.  The
$l\geq2$ waves and the high energy are input.
Although RE are valid
for $\sqrt{s}\leq8M_\pi\simeq1120\,$MeV, for now
we have implemented them only up to $\sqrt{s}\simeq2M_K$.

The aim of our group when using RE combined with FDR
has been to improve the precision of scattering data analysis, to test ChPT,
and, still in progress, to obtain precise determinations of the $\sigma$ 
resonance.
The Bern group \cite{Bern} has also carried out a series of RE 
analysis, with and without ChPT constraints, using as input 
phenomenological parametrizations
for the $l\geq2$ waves and above 800 MeV for the other waves,
 as well as some Regge input. When using ChPT constraints,
they find, $a^{(0)}_{0}=0.220\pm0.005\,M_\pi^{-1}$ 
and $a^{(2)}_{0}=-0.0444\pm0.0010\,M_\pi^{-1}$, an extremely precise claim,
together with predictions for other scattering lengths and 
the S and P wave phase shifts up to 800 MeV. Although of 
the original input, particularly the Regge theory and the 
D waves, was questionable \cite{critica}, it certainly seems to have
a very small influence in the threshold region 
of the scalar waves \cite{Caprini:2003ta}.
In addition, the Krakow-Paris \cite{Kaminski:2002pe} and Paris 
\cite{DescotesGenon:2001tn} groups have performed other RE analysis. 
The former resolved a long-standing ambiguity, discarding the 
so-called "up"' solution, including in their analysis a study
using polarized target data. The latter checked the calculation in \cite{Bern}
and claimed an small discrepancy in the Olsson sum rule. 

\section{Sketch of our analysis}
\vspace*{-.3cm}
The approach we have followed over a series of works \cite{Kaminski:2006qe} 
can be sketched as follows: 

(1)
We obtain simple ``Unconstrained Fits to Data`` (UFD)
 to each $\pi\pi$ scattering wave so that it
can be improved independently if needed.
High energy fits use Regge theory. 
As our precision improves, we refine our
 fits with more flexible parametrizations or with
new and more precise data, as, for example,
in \cite{Yndurain:2007qm} 
where we included the newest and very precise $K_{l_4}$ data \cite{Batley:2007zz}.

(2) We then check how well these UFD satisfy FDRs and RE and some sum rules. 
To our surprise some of the most widely used data parametrizations 
fail to satisfy FDRs or the sum rules. 
Thus we choose the data parametrizations 
in better agreement with FDRs. 

(3) Our best results are obtained by 
imposing dispersion relations (both FDRs and RE) and some crossing
sum rules (SR)
on the data fits, now called "`Constrained Fits to Data"' (CFD),
which are thus consistent
with analyticity, unitarity, crossing, etc...

A dispersion relation $i$ is well satisfied at a point $s_n$
if the difference $\Delta_i$ between the left and right 
sides in Eqs.(\ref{FDR1}),(\ref{FDR2}) and (\ref{Roy}) is small
relative to its uncertainty $\delta\Delta_i$. 
Thus, when 
the average discrepancy (similar to an averaged $\chi^2/(d.o.f)$)
\begin{equation}
\bar{d}_i^2\equiv\frac{1}{\hbox{number of points}}
\sum_n\left(\frac{\Delta_i(s_n)}{\delta\Delta_i(s_n)}\right)^2\leq1
\label{avdiscrep}
\end{equation}
it implies that the corresponding dispersion relation is satisfied within uncertainties.
 In practice, the values of $s_n$ are taken at intervals of 25 MeV. 
This measure is also used \cite{Kaminski:2006qe} to obtain the Constrained Fit to Data, 
by minimizing:
\begin{equation}
\chi^2\equiv
W\sum_i \bar{d}_i^2
+\bar{d}^2_I+\bar{d}^2_J+\sum_k\left(\frac{p_k-p_k^{\rm exp}}{\delta p_k}\right)^2,
\end{equation}
where $p_k^{exp}$ are all the parameters of the different UFD parametrization for each wave or Regge trajectory, 
thus ensuring the data description, and
$d_I$ and $d_J$ are the discrepancies for a couple of crossing sum rules. The weight $W=9$ was estimated from
the typical number of degrees of freedom needed to describe the shape of the amplitude.

The result of this program is, on the one hand, a set of precise Constrained
Fits to Data,  that satisfy very well 
all dispersion relations within uncertainties. 
Remarkably, all the waves are given in terms
of very  simple parametrizations that can be found in \cite{Kaminski:2006qe} which are
very easy to use for phenomenological purposes. On the other hand, 
we have 
the outcome of Roy Eqs. themselves that allows us to extend
the calculation to the complex plane and look for poles associated to resonances and study their parameters.

In particular the best determination 
of threshold parameters is obtained by using the CFD 
set directly or inside appropriate sum rules \cite{Kaminski:2006qe}. 
For the S0 and S2 waves, we find: $a^{(0)}_{0}=0.223\pm0.009\,M_\pi^{-1}$ 
and $a^{(2)}_{0}=-0.0444\pm0.0045\,M_\pi^{-1}$, in remarkable agreement
with the predictions in \cite{Bern}, 
that extends also to the $P$ wave scattering length.
There are however, some disagreements of 2 to 3 standard deviations 
in the P-wave slope, and also in some $D$ wave parameters. 
In general, for the controversial S0 wave,
the agreement is fairly good only up to roughly 450 MeV,
but from that energy up to 800 MeV those predictions 
deviate slightly from our data analysis.
This means that
we find from our Roy Eqs. and FDR data analysis 
a somewhat different pole from that of the Bern group (see R. Kaminski talk
in this conference).
Let us emphasize that we are talking about a deviation of a few degrees 
that only affects the sigma mass and width determination
 by ten or twenty MeV at most, 
which is a remarkable improvement compared with the situation 
just a few years ago and the
huge and extremely conservative uncertainties of hundreds of MeV
quoted in the PDG at present \cite{Amsler:2008zz}.

\subsection{Latest developments}
 
 In this conference we report preliminary results on two issues:
 First, we have recently implemented once-subtracted Roy Eqs,
(denoted GKPY for brevity)\cite{Kaminski:2008fu}.
This is motivated by the interest
 in the $450 {\rm MeV}<\sqrt{s}<2 M_K $ region of the S0 wave, 
dominated by the $\sigma$-resonance.
 The large low-energy experimental uncertainties in the 
S2-wave translate into large uncertainties for the dispersive output for the
S0 wave in 
that region when using the standard twice subtracted Roy Eqs. (shaded area in top panel of Fig 1),
 but not for the once subtracted ones, 
which have a much smaller uncertainty (Fig 1, Bottom).

\begin{figure}[t]
\begin{center}
\vspace*{-20pt}
\includegraphics[angle=-90,width=7cm]{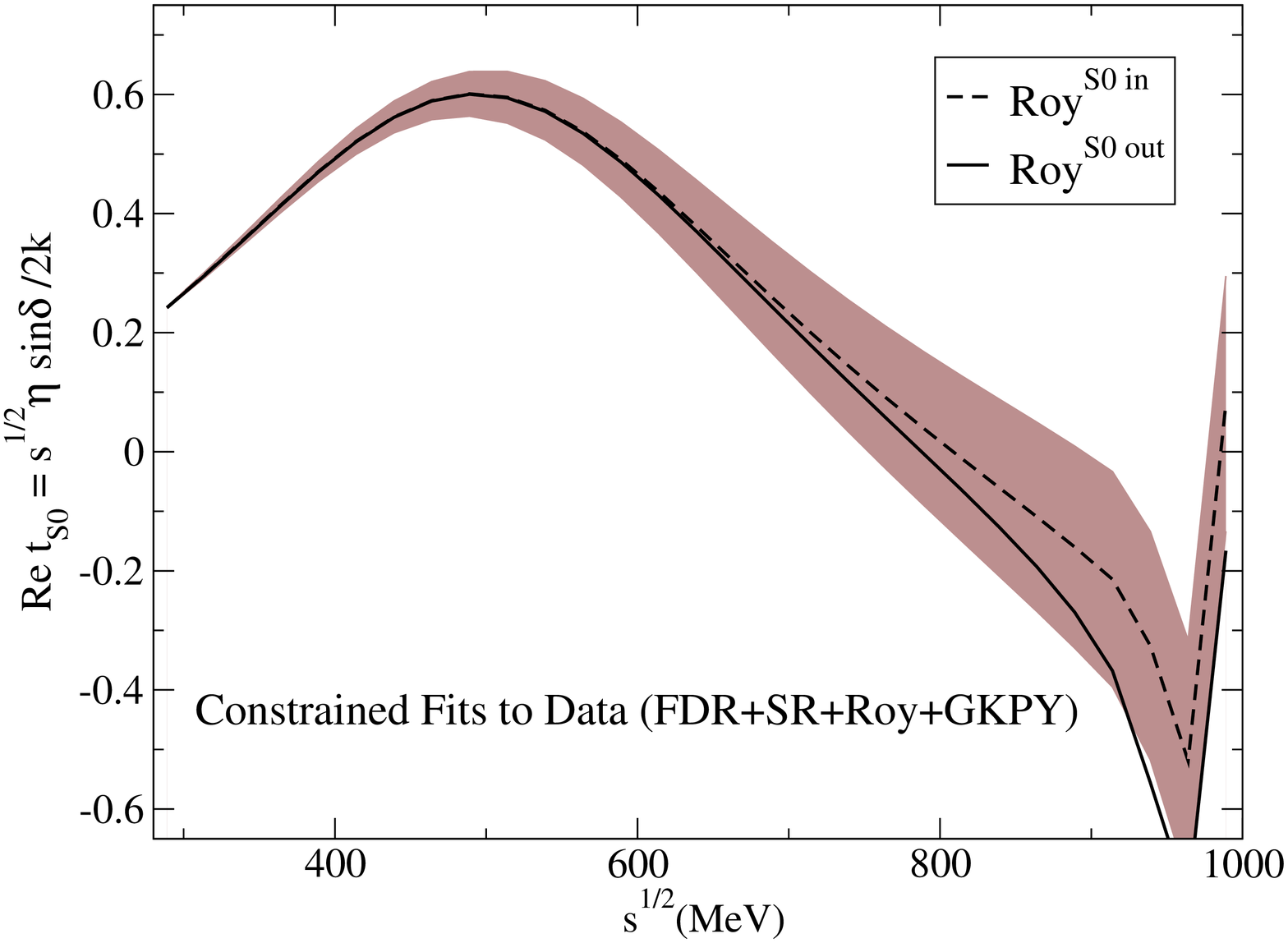}

\vspace{-15pt}
\includegraphics[angle=-90,width=7cm]{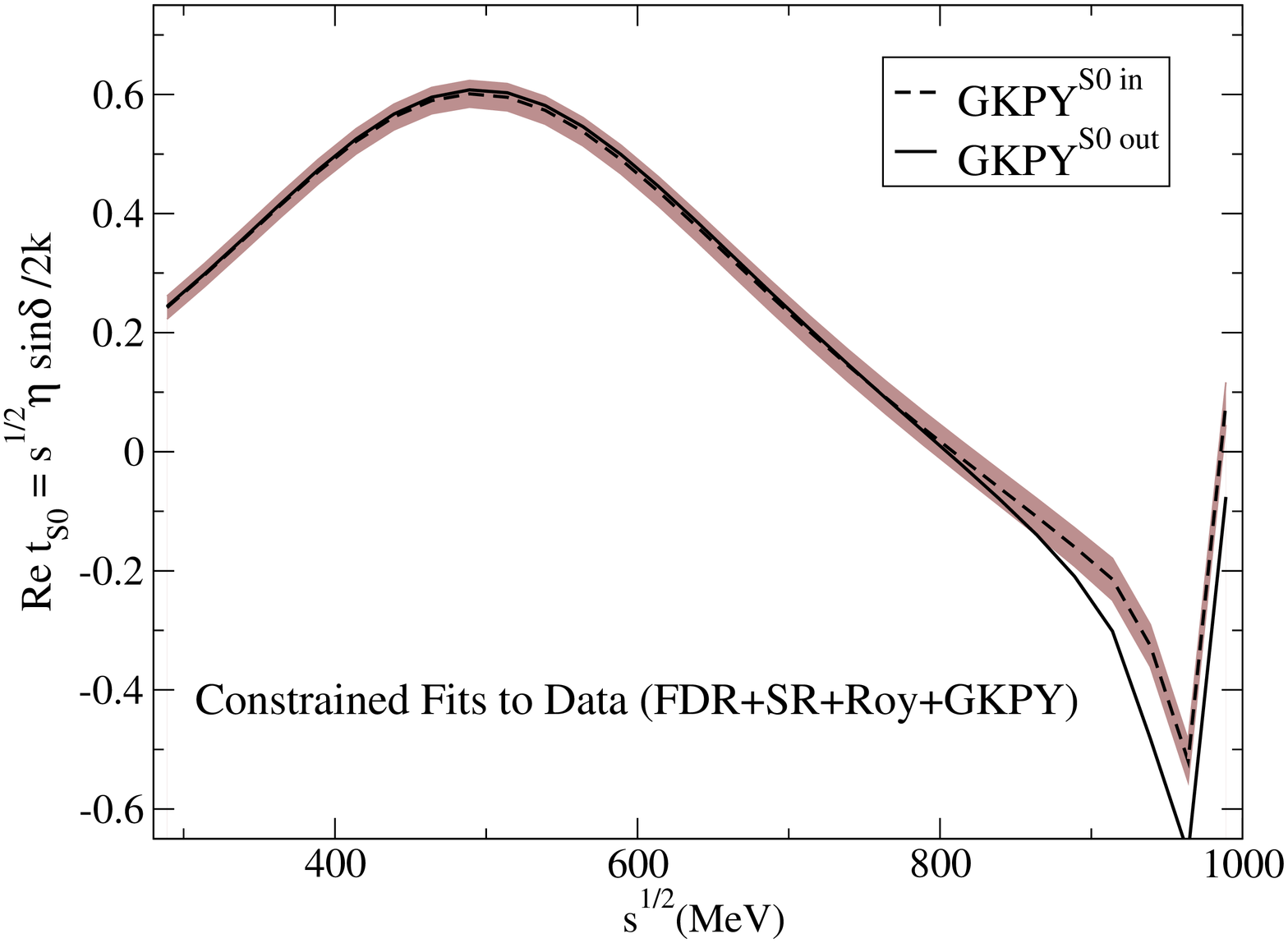}
\end{center}
{\small Figure 1.
Above 500 MeV, the uncertainty in standard S0 Roy Eqs. (Top)
is much larger than with once subtracted 
Roy Eqs. (Bottom).}
\vspace*{-20pt}
\end{figure}

 Second, we have improved the matching
between the S0 wave parametrizations at low and intermediate energies,
which occurs at 932 MeV, imposing also continuity in the 
first derivative. This follows a suggestion
\cite{Leutwyler:2006qp} to explain the roughly $2\sigma$ 
level discrepancies in the S0 wave
 between our analysis and that of the Bern group in the $400-900$MeV region. 
 In Fig.2 we compare the CFD in \cite{Kaminski:2006qe} (dashed line)
and the new one with the improved matching (continuous line),
which only differ a little above 932 MeV. 
Note that, for clarity, 
we do not
provide data points, which are nevertheless 
reasonably well described by both parametrizations 
when taking into account experimental errors.
Hence, the disagreement is {\it not caused} by a poor matching,
but we are studying further suggestions in \cite{Leutwyler:2006qp}. 
Nevertheless this better matching 
improves the Roy and GKPY Eqs. fulfillment
above 850 MeV, as seen by comparing the bottom panels in Figs.1 and 2.
 
\begin{figure}[t]
\begin{center}
\vspace*{-10pt}
\includegraphics[angle=-90,width=6.3cm]{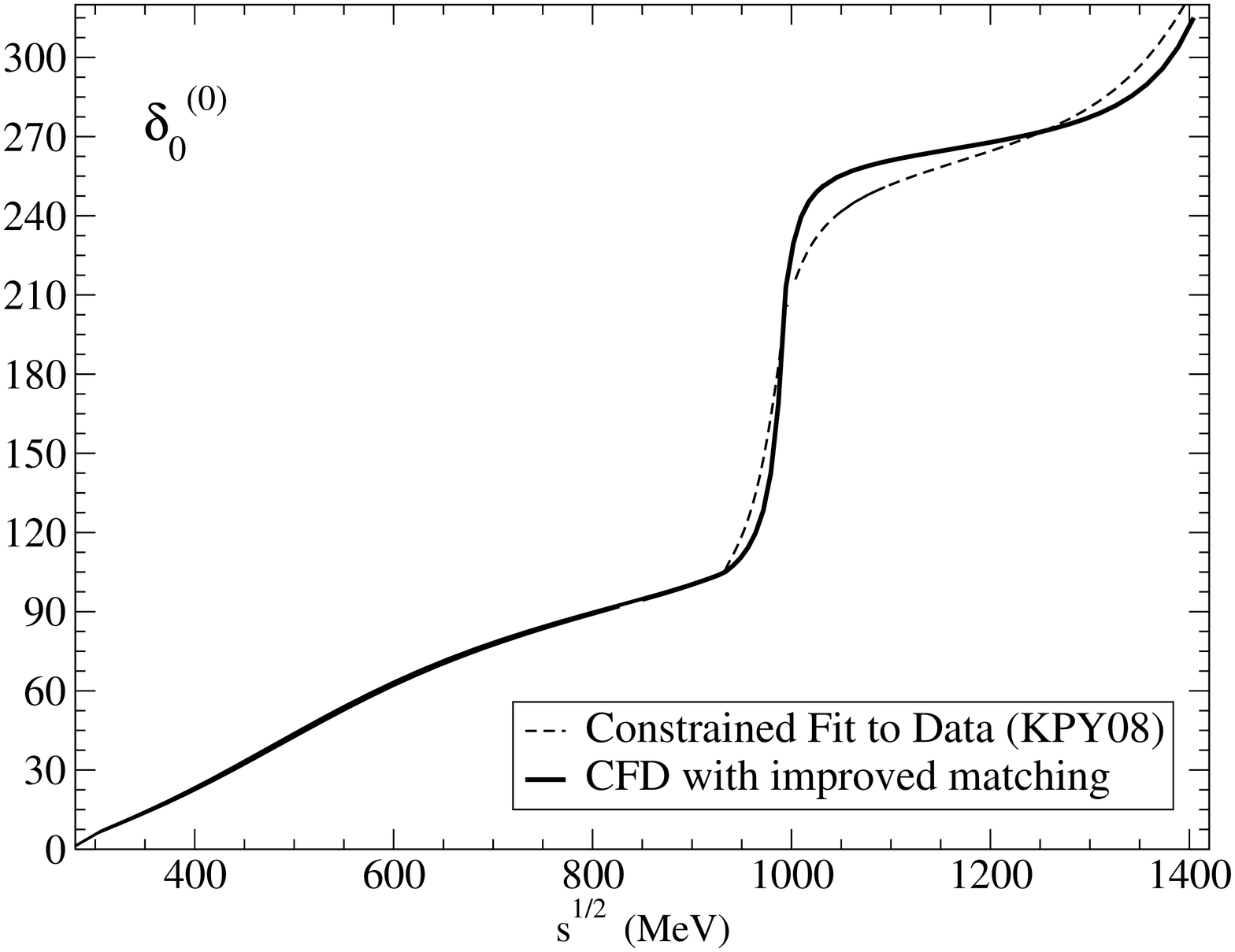}

\vspace*{-4pt}
\includegraphics[angle=-90,width=7cm]{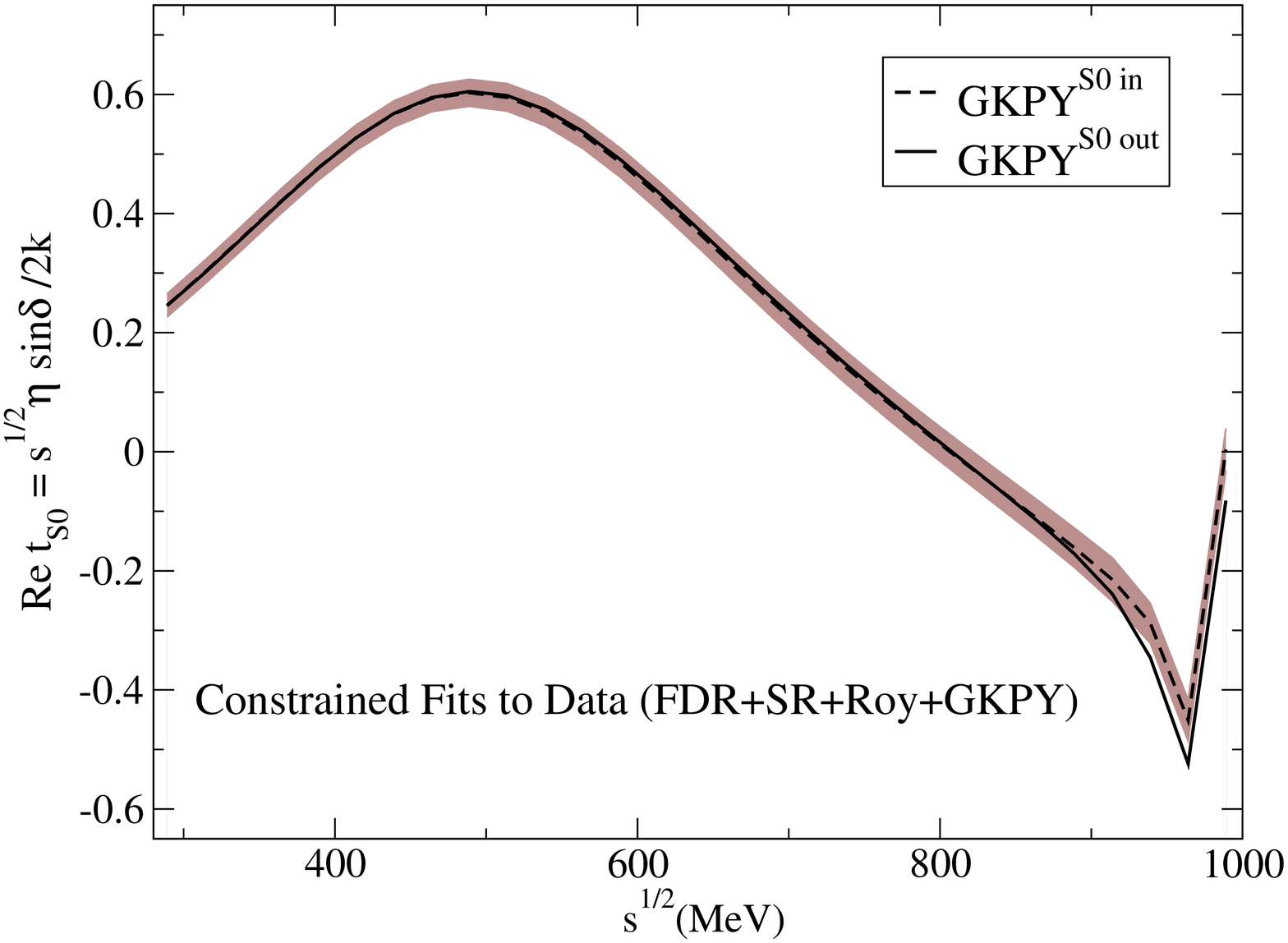}
\end{center}
{\small Figure 2.
Top: The improved matching only changes slightly
 the  S0 wave above 932 MeV. Bottom: The agreement 
GKPY Eqs. is slightly improved above 850 MeV (compare with Fig.1 Bottom). }
\vspace*{-20pt}
\end{figure}

Finally, in order to show how well our preliminary
CFD set satisfies the nine dispersion relations,
we list in
Table 1 the average discrepancies, defined in Eq.(\ref{avdiscrep}), 
for the three FDR, up to two different energies, 
and for the three Roy Eqs. up to $\sim2M_K$,
either with the standard two subtractions or only with one. 
From that Table it is clear that this CFD set satisfies remarkably 
well all dispersion relations within uncertainties

In summary the CFD set provides a 
model independent and very precise description
of the $\pi\pi$ scattering data consistent with analyticity and crossing
that can be easily used for phenomenological purposes. Work is in progress
to obtain an even more accurate description for threshold
parameters than the one we had, and (see R. Kaminski talk)
by using it inside once-subtracted Roy Eqs.,
to obtain a precise determination from data of the $\sigma$ pole parameters.

\section*{Acknowledgments}
J.R.P. thanks A. Galindo, A. Gonz\'alez Arroyo and  M.J. Herrero, for kindly providing
information about Paco Yndur\'ain's career and H. Leutwyler for his comments about the matching of the S0 wave.

\begin{table}
\begin{tabular}{|l|c|c|}
\hline
FDRs
&{\small $s^{1/2}\leq 932\,$MeV}&{\small $s^{1/2}\leq 1420\,$MeV}\\
\hline
$\pi^0\pi^0$& 0.13 & 0.31\\
$\pi^+\pi^0$& 0.83 & 0.85\\
$I_{t=1}$& 0.13 & 0.70\\
\hline
\hline
&\multicolumn{2}{c|}{$s^{1/2}\leq 992\,$MeV}\\
Roy Eqs.& 2 subs. & 1 sub. (GKPY)\\
\hline
S0& 0.06&0.49\\
S2& 0.13&0.11\\
P& 0.11& 0.23\\
\hline
\end{tabular}
{\small Table 1. Average discrepancies $\bar d^2$ of the 
Constrained Fit to Data  for each 
forward dispersion relation.}
\vspace{-20pt}
\end{table}

\end{document}